\documentclass[aps,jcp,twocolumn]{revtex4-1}
\usepackage{amsmath}
\usepackage{amsfonts}
\usepackage{graphicx}
\usepackage{epstopdf}
\usepackage{amsthm}

\makeatletter
\DeclareFontFamily{OMX}{MnSymbolE}{}
\DeclareSymbolFont{MnLargeSymbols}{OMX}{MnSymbolE}{m}{n}
\SetSymbolFont{MnLargeSymbols}{bold}{OMX}{MnSymbolE}{b}{n}
\DeclareFontShape{OMX}{MnSymbolE}{m}{n}{
    <-6>  MnSymbolE5
   <6-7>  MnSymbolE6
   <7-8>  MnSymbolE7
   <8-9>  MnSymbolE8
   <9-10> MnSymbolE9
  <10-12> MnSymbolE10
  <12->   MnSymbolE12
}{}
\DeclareFontShape{OMX}{MnSymbolE}{b}{n}{
    <-6>  MnSymbolE-Bold5
   <6-7>  MnSymbolE-Bold6
   <7-8>  MnSymbolE-Bold7
   <8-9>  MnSymbolE-Bold8
   <9-10> MnSymbolE-Bold9
  <10-12> MnSymbolE-Bold10
  <12->   MnSymbolE-Bold12
}{}

\let\llangle\@undefined
\let\rrangle\@undefined
\DeclareMathDelimiter{\llangle}{\mathopen}%
                     {MnLargeSymbols}{'164}{MnLargeSymbols}{'164}
\DeclareMathDelimiter{\rrangle}{\mathclose}%
                     {MnLargeSymbols}{'171}{MnLargeSymbols}{'171}
\makeatother

\newcommand{\inner}[2]{\left< #1 \vphantom{#2} ,
  #2 \vphantom{#1} \right>} 
\newcommand{\braket}[3]{\left< #1 \vphantom{#2#3} ,
 #2  #3 \vphantom{#1#2} \right>} 

\newcommand*{\avg}[1]{\left\llangle #1 \right\rrangle}

\newcommand*{\erf}{\operatorname{erf}}
\newcommand*{\erfc}{\operatorname{erfc}}

\begin{document}

\title{ Variational Formulation of Local Molecular Field Theory} 

\author{ David M. Rogers}
\affiliation{ National Center for Computational Sciences, Oak Ridge National Laboratory, Oak Ridge, TN}
\email{ rogersdm@ornl.gov}

\begin{abstract}
  In this note, we show that the Local Molecular Field theory of Weeks et. al. can be re-derived as
an extremum problem for an approximate Helmholtz free energy.
Using the resulting free energy as a classical, fluid density functional yields an implicit
solvent method identical in form to the Molecular Density Functional theory
of Borgis et. al., but with an explicit formula for the `ideal' free energy term.
This new expression for the ideal free energy term can be computed from
all-atom molecular dynamics of a solvent with only short-range interactions.
The key hypothesis required to make the theory valid is
that all smooth (and hence long-range) energy functions
obey Gaussian statistics.
This is essentially a random phase approximation for perturbations
from a short-range only, `reference,' fluid.
This single hypothesis is enough to prove that the self-consistent LMF procedure
minimizes a novel density functional whose `ideal' free energy is
the molecular system under a specific, reference Hamiltonian, as opposed to the non-interacting gas
of conventional density functionals.  Implementation of this new
functional into existing software should be straightforward and robust.
\end{abstract}

\date{\today}
\maketitle

  Classical fluid density functional theories are intimately connected to the Ornstein-Zernike
theory relating pair correlation functions to the likelihood of density fluctuations.
In 1964, Percus and Yevick assumed the direct correlation function was zero
between hard spheres that were not in contact, and derived radial distribution functions
and equations of state for the hard sphere fluid that are still useful today.\cite{davis,storq18,athor18}
Subsequent work attempted to find appropriate hypotheses for the direct correlation
function in fluids with long-range interactions.  The mean spherical approximation (MSA)
is one of the simplest such closures, and assumes $c(r) = -\beta u_{LR}(r)$ outside contact.\cite{mwert71,wmadd80}
However, it fails to adequately describe the structure or equation of state for
the dipolar hard-sphere fluid.\cite{jbark76}

  In 1973, Chandler argued for the viewpoint that closure relations (relating $c$ and $g$ using the
pairwise energy, $u$) are not as primary as density functionals (relating $g$ to the fluid free energy).\cite{dchan73}
Every density functional produces a thermodynamically consistent model.
Further, an `exact' functional exists for every fluid that computes the free energy
required to produce a given density distribution.
The equilibrium density was then found as a minimum the density functional.
For a mathematical exposition of this theory, see Ref.~\citenum{eschrig}.
For modern developments tracing their origin to those works, see Refs.~\citenum{svolo12,jjohn16}.
This re-cast the impossible problem of divining a closure relation with the difficult problem of
finding an appropriate density functional.

  In 1989, Rosen showed that the Percus-Yevick equation of state could be derived
from a local density functional using `fundamental measures,' which describe local densities
of particles, contact surfaces, and excluded volumes.\cite{yrose89}
These fundamental measures were also found to help simplify some expressions for
the MSA description of ionic solutions.\cite{lblum91}
For a more complete review and recent applications of those works,
see Refs.~\citenum{rroth10,hyin19}.  An advanced computational implementation is
available in the Tramonto project.\cite{msear03,mhero07}
The many success of this approach for fluids with only short-range interactions
have demonstrated that structure and solvation free energies be described
well by integral equation methods.

  Nevertheless, fluid density functional theories are known to poorly describe fluids
that contain long-range interaction energies (e.g. charge-charge or dipole-dipole).
Even long-range dispersion is challenging.  In 1980, Weeks, Chandler,
and Anderson derived a Lennard-Jones equation
of state as a first-order perturbative correction to the hard-sphere fluid.\cite{wca}  Verlet and
Weis argued that the radial distribution function should be corrected as well,\cite{lverl72}
and noted that the first-order correction procedure must break down at (intermediate?)
densities.  The straightforward use of perturbation theory to add pairwise interactions
to fluid density functional theory leads to functionals that are a sum of local, hard-sphere-like
free energies, and quadratic, density-density interaction energies:\cite{revan92,mleve12}
\begin{equation}
F^\text{pert}[\rho] = F^\text{id}[\rho] + \frac{1}{2} \braket{\rho - \rho_0}{u^\text{LR,eff}}{(\rho - \rho_0)}
\label{e:pert}
\end{equation}
It is not hard to connect this to the Ornstein-Zernike theory and find that $u^\text{LR,eff}$
should be proportional to the fluid's direct correlation function.\cite{pdt6}

  Equation~\ref{e:pert}, however, is known to give poor results because the direct
correlation function does not remain constant as the pairwise interaction is increased.
This has lead to a re-examination of the original closure relations, which state
that, technically, $\beta u^\text{LR,eff} = -\int_0^1 d\lambda (1-\lambda) c_\lambda$,
where $c_\lambda$ is the direct correlation function when the solute and solvent
interact via the scaled-down potential, $\lambda \Delta U$.
An alternative viewpoint on the same problem of steric overlaps
is that there is a nonzero probability for very unfavorable solute-solvent interaction
energies in a single step particle insertion.\cite{droge08,dkilb18}

  Rather than taking this track, Widom noted that if solvation is split up into a local
cavity formation step plus a subsequent, long-range interaction step, then
the local step can be estimated from the hard-sphere equation of state and the long-range
part can be estimated by 1-step perturbation.\cite{bwido82}  Taking this idea into the modern, density
functional viewpoint, the long-range step is very well approximated by linear response.
Weeks and co-workers have explored this linear-response regime
and reported that, as expected, it is associated with
Gaussian fluctuations in the long-range potential.\cite{ychen06,jrodg08,rrems16}
Other workers have cautioned that fluctuations in the potential
may be approximately Gaussian, but their widths depend
on the geometry of the solute.\cite{tpoll18}

  In this work, we derive a new density functional theory based on the single
hypothesis that all smooth (and hence long-range) energy functions
obey Gaussian statistics.  This single hypothesis is enough to prove that
the self-consistent LMF procedure minimizes a novel density functional
whose `ideal' free energy is a short-range only fluid, as opposed to the non-interacting gas
of conventional density functionals.  Further, this density functional
has the same form as the molecular density functional theory investigated
by Borgis and co-workers.\cite{rrami02,mleve12,gjean13,gjean16}
The latter finding is surprising, since
their solvation free energy functional was originally thought to be an extension
of the hypernetted chain closure.



  Let us consider the problem of computing the following
free energy,
\begin{equation}
\beta A(G, \phi) = -\ln \int d\nu_0 \; e^{-\tfrac{1}{2}\braket{\hat n}{G}{\hat n} - \inner{\hat n}{\phi}}
,\label{e:A}
\end{equation}
where $d\nu_0$ is a probability distribution over molecular coordinates
and $\hat n(r) = \sum_{i} \delta(r - \hat r_i)$ is the local density of molecules
in configuration $\{r_i\}$.
Carets denote random variables that are functions of the microstate.
This free energy provides a large deviation function for the density, $\avg{\hat n}$,
and its fluctuations, $\avg{ \hat n \hat n^T }$.


  For example, two neighboring boxes containing
$n$ and $m$ gas molecules and following independent Poisson
distributions with mean $\lambda$ would have
\begin{equation}
\begin{split}
A_2(G,\phi) &= 2 \lambda - \ln \sum_{n,m=0} \frac{ \lambda^{n+m}}{n! m!}
    \exp \Big(  -\phi_1 n - \phi_2 m \\
& - \frac{G_{11} n^2 + 2 G_{12} n m + G_{22} m^2}{2} \Big)
.
\end{split}
\end{equation}
Large deviation theory then states that the distribution over $n,m$
is asymptotically the exponent of the entropy,
\begin{equation}
S_2(G,n,m) = \inf_\phi \left [ \phi_1 n + \phi_2 m - \beta A_2(G,\phi) \right]
.
\end{equation}
When $n,m$ are both large, then the Poisson distributions approach
a Gaussian, and we can provide approximate expressions for both of
the above.  More to the point, Gaussian fluctuations about the mean
provide a set of relations between $\rho$, $\phi$, $A$ and $S$
as $G$ is varied.

  Applying the analogous procedure to approximate $\hat n$
by a Gaussian distribution for $d\nu_0$
with mean and variance $\rho_0$ and $\Sigma_0$,
we find the Gaussian approximation to Eq.~\ref{e:A},
\begin{equation}
\begin{split}
A(G, \phi) \approx \frac{1}{2} \Big[ & \ln |\Sigma_0 G + I|
+ \braket{ \rho_0 }{\Sigma_0^{-1}}{ \rho_0 } \\
&- \inner{ \rho}{ \Sigma_0^{-1}\rho_0 - \phi }
\Big]
\end{split}
. \label{e:gaus}
\end{equation}
The density and variance transform as,
\begin{equation}
\rho = (\Sigma_0 G + I)^{-1} (\rho_0 - \Sigma_0 \phi),
\quad
\Sigma^{-1} = \Sigma_0^{-1} + G \label{e:path}
.
\end{equation}
Implicit differentiation of the former equation gives
$d\phi = -dG \rho$ along a path with constant density.
Integration along this path could have been used to
carefully avoid the full Gaussian integral of Eq.~\ref{e:gaus}.



\section{ Solvation Free Energy}

  To find a correspondence with LMF, take the reference system
 to be a short-range fluid,
 \begin{equation}
 d\nu^0 = \frac{ e^{-\beta \hat U_\text{SR}} dx }{ \int dx \; e^{-\beta \hat U_\text{SR}} }
 .
 \end{equation}
 while the full potential energy function of the fluid is
 $\hat U = \hat U_\text{SR} + \frac{\beta^{-1}}{2} \braket{\hat n}{G}{\hat n}$.
 Then the solvation free energy of a solute which interacts with
every fluid particle via the potential $\phi(r)$, is
\begin{equation}
\mu^\text{ex}(\phi) = A(G, \phi) - A(G, 0). \label{e:solv}
\end{equation}

  The solvation free energy in the reference fluid
could be found from
\begin{equation}
\mu^\text{ex}_\text{ref}(\phi_1) = A(0, \phi_1) - A(0, \phi_0)
,
\end{equation}
where $\phi_1 = \phi + G\rho$ and $\phi_0 = G \rho'$
are potentials which set the reference
fluid density to the corresponding real fluid density
at states $(G,\phi)$ and $(G,0)$, respectively.

  Taking the solvation free energy of Eq.~\ref{e:solv} as our goal,
we can compute it from the reference system with a thermodynamic
cycle: $(G,0) \to (0,\phi_0) \to (0,\phi_1) \to (G,\phi)$.
The Gaussian approximation provides a clean answer for
\begin{equation}
\Delta \Delta A = \mu^\text{ex}(\phi) - \mu^\text{ex}_\text{ref}(\phi_1)
= -\frac{1}{2} \left[ \braket{\rho}{G}{\rho} - \inner{\phi_0}{\rho'} \right]
.
\end{equation}

  Finally, we can state the LMF free energy functional,
\begin{equation}
\beta \mu^\text{ex}(\phi) = \inf_x \Big[
\beta \mu^\text{ex}_\text{ref}(x) + \inner{\rho}{\phi-x} + \frac{1}{2} \braket{\rho}{G}{\rho}
\Big] -  \frac{1}{2} \inner{\phi_0}{\rho'} 
.\label{e:muex}
\end{equation}
Provided we always evaluate at points where $\rho(x) = \partial \beta \mu^\text{ex}_\text{ref} / \partial x$.
Taking the variations indicated shows that $x = \phi + G\rho$ is the self-consistent
potential satisfying $\rho = \rho(0, x)$.  Note that the minimum above is unique
for our Gaussian case because the second variation of the right-hand side
is $(\Sigma_0 G + I) \Sigma_0$, which we assumed to be positive-definite
in writing Eq.~\ref{e:path}
(because otherwise the fluctuations would be negative under pairwise potential $G$).
Also, the potential function minimization used in Eq.~\ref{e:muex}
is stable in cases where $G \to 0$ and usually has a larger
domain than the corresponding formulation in terms of $\rho$.

\section{ Connection to Molecular Density Functional Theory}

  To connect to conventional molecular density functional theory, we need the free energy
for an ideal gas,
\begin{equation}
\beta \mu^\text{ex}_\text{id}(x) = -\ln \sum_{n=0}^\infty \frac{e^{-nx-\lambda}\lambda^n}{n!} = \lambda (1 - e^{-x})
.
\end{equation}
Contrary to popular nomenclature, standard molecular density functional theory is not based on this
free energy, but instead its Legendre transform $\inner{\rho}{x}$
\begin{equation}
-s^\text{ex}_\text{id}(\rho) = \beta \mu^\text{ex}_\text{id}(x) - \rho x = \rho \ln\frac{\rho}{\lambda} - \rho + \lambda
.
\end{equation}
Here $\rho = \partial \beta \mu^\text{ex}_\text{id} /\partial x$.
This explains the composite expression for the solvation free energy
in HNC / MDFT using
\begin{equation}
s^\text{ex}_\text{mdft}(\rho) \equiv s^\text{ex}_\text{id}(\rho) + \frac{1}{2} \braket{\rho-\rho'}{C}{(\rho-\rho')}
,
\end{equation}
where $C$ is the direct correlation function mentioned in connection with Eq.~\ref{e:pert}.
Legendre-transforming to the excess entropy provides
\begin{equation}
\beta\mu^\text{ex}_\text{mdft}(\phi) = \inf_\rho \left[\inner{\rho}{\phi} - s^\text{ex}_\text{mdft}(\rho) \right]
.
\end{equation}

  To compare, note that Eq.~\ref{e:muex} is identical when phrased in terms of $\rho$ rather than $x$
as long as $\delta \beta\mu^\text{ex}_\text{ref} / \delta x = \rho$ is maintained.
Thus, the excess entropy implicit in Eq.~\ref{e:muex} is,
\begin{equation}
s^\text{ex}(\rho) = s^\text{ex}_\text{ref}(\rho) - \frac{1}{2} \braket{\rho-\rho'}{G}{(\rho-\rho')} - \braket{\rho}{G}{\rho'}
.
\end{equation}

\begin{figure*}
\begin{tabular}{ll}
\includegraphics[width=4.75in]{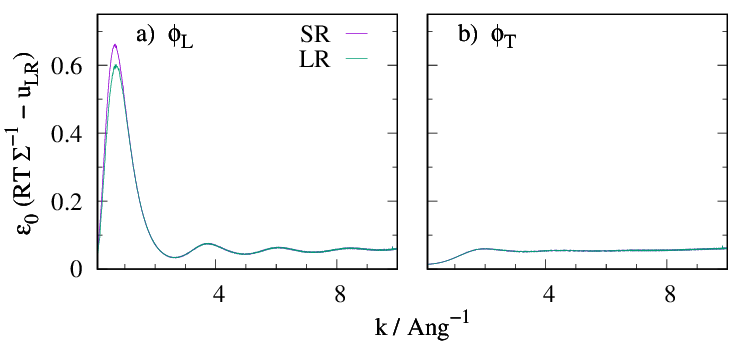} &
\hspace{-2.5em} \includegraphics[width=2.65in]{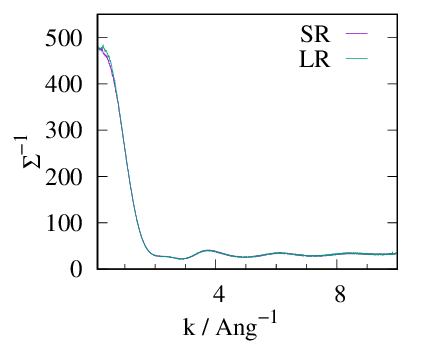}
\end{tabular}
\caption{Comparison of the inverse correlation functions of SPC/E water
with and without long-range $(\erf(r \sigma)/r)$ electrostatics included.}\label{f:compare}
\end{figure*}

  The two formalisms can be compared by checking derivatives of the entropy functional,
$s(\rho)$.  The first derivative provides the external potential required to set a given density,
and the second derivative gives the negative of the inverse correlation function
at a given density.
\begin{align}
\phi_\text{mdft}(\rho) &= -\ln\frac{\rho}{\lambda} + C (\rho - \rho'),
  & \phi(\rho) &= \phi_\text{ref}(\rho) - G \rho \\
\Sigma_\text{mdft}^{-1}(\rho) &= \frac{\delta(r - r')}{\rho(r)} - C,
  & \Sigma^{-1}(\rho) &= \Sigma^{-1}_\text{ref}(\rho) + G
  .
\end{align}

  These relations prove the simple interpretation
that MDFT is a quadratic perturbation theory from a non-interacting
ideal gas state and that LMF is a quadratic perturbation theory
from a short-range, reference state.
This goal seems to be what was aimed for in Ref.~\citenum{jperc11}.

\section{ Test of Hypotheses }

  The key hypotheses leading to Eq.~\ref{e:path} was that
long-range interactions add to the {\em inverse} correlation function.
If this is true, then the long-range pair energy is enough to predict
the long-range correlations.
It causes us to shift our focus away from $C$ (or OZ closure relations)
and toward understanding solvation in the reference system.

  Testing this hypothesis is a simple matter of calculating the
correlation functions, $\Sigma(\rho, G)$, as a function of
the long-range pair energy, $G$.  Fig.~\ref{f:compare} shows
the result of this calculation, carried out in Fourier space using the method of
Ref.~\citenum{droge18}.  A periodic cell of 8000 SPC/E water
molecules was simulated with LAMMPS for 30 ns using SHAKE
constraints and a 2 fs timestep.  Long-range interactions
were handled using PPPM with an Ewald distance
splitting of $\sigma = 4$\AA{}.  The simulation labeled SR-only
did not include PPPM's k-space summation, so that
the electrostatic potential energy was truncated to,
\begin{equation}
E^\text{Coul}_{SR} = \frac{1}{2} \sum_{i\ne j} \frac{ q_i q_j \erfc(\sigma r_{ij})}{ 4\pi\epsilon_0 r_{ij}}
.
\end{equation}

  Our result provides a more detailed confirmation that the density fluctuations
in long-range interacting liquids can be predicted from corresponding short-range fluids.

  The analogous comparison for the hypernetted chain closure form of MDFT would
plot the direct correlation function, $C$, as all pairwise interactions are scaled from zero
to their final value.  Obviously, this is a much larger perturbation.  The principle
difficulty with predicting $C$ during this switching process is caused
by short-range interactions, which strongly re-structure the fluid.  In contrast, the
LMF perturbation changes the fluid density over larger length-scales.
Its largest difficulty is the prediction of the properties of the reference fluid.

\section{ Conclusions}

  The intense effort spent on liquid-state density functional theories
in the 1980s focused on the Ornstein-Zernike equation, and was
able to derive useful functional forms for radial distribution functions
based on known long range behavior of the direct correlation function.
However, it required making assumptions about closures that resulted
in conflicting, non-systematic ways to extract short range
structure and thermodynamic functions.\cite{hyin19}
The fluid density functional theories that succeeded them
directly addressed thermodynamic functions, but had
similar problems ``bridging" condensed and ideal gas
phases.  From the perspective of this work, starting from
an ideal gas free energy expression seems to have been the
incorrect assumption hindering success of this theory.


  This work has confirmed that the Gaussian the fluctuation assumption works
well when applied separately to the long-range structure.
Based on this, it presented a split-range density functional
theory that uses the short-range fluid as a reference state
instead of an ideal gas.
Our explicit identification of a reference system should be helpful
for cross-over studies with quantum density functional theories.\cite{jtolo04}
The connection to the random phase approximation also suggests
a quasiparticle approach to kinetics.\cite{dbohm53}

  The results of the present work should be applied to more large-scale
tests of solvation, including computing phase-diagrams of complicated
solvents and calculating surface forces governing macromolecular
adhesion.\cite{gdayh19}

  Parameterization of the short-range density functional will become
increasingly valuable for MDFT studies using this approach.
This goal seems obvious from a consideration of the intense effort
invested into the parallel context of the electron gas in quantum DFT.\cite{wkohn99}
Because the interactions have been made short-ranged,
this parameterization could be done with the help of finite-sized
simulations.  Larger choices for the range separation distance ($\sigma$)
will make the Gaussian assumption more accurate at the expense
of increased difficulty during this parameterization step.

  Despite the promises of this theory, there are also several ares for concern.
Water has a strong quadrupolar interaction, responsible for a
contribution to solvation free energies for ions.  Water's quadrupolar
moment is responsible for the controversy over the potential drop
at a water/vacuum interface.\cite{kleun10,skath11}
Because quadrupole-ion interactions scale as $r^{-3}$,
they could cause significant energetic and structural changes
extending beyond $\sigma$.

  A key objective for future work is the development of numerical
methods that can carry out the formalism for electrolytes and non-aqueous solvents.
This work is straightforward, with robust implementations of solution
density functional theory already available.\cite{msear03,mhero07}
Incorporation of the present model into those methods
has not yet been attempted because of the necessity of switching
to a potential-function basis and representing solvent densities as
multicomponent vectors (containing positional and orientational densities).

\section*{Acknowledgments}

I thank the USF research foundation for its partial support of this work.

\bibliographystyle{unsrt}

\end{document}